\documentclass[11pt,a4paper]{iopart}
\pdfoutput=1
\usepackage[T1]{fontenc}\usepackage[latin1]{inputenc}
\usepackage{graphicx,color,booktabs,wrapfig,microtype,sidecap,floatflt,afterpage,cite,hvfloat} 
\usepackage[charter]{mathdesign}
\renewcommand{\figurename}{Figure}

\newcommand{\half}{\frac{1}{\protect\raisebox{0.8pt}{\scriptsize 2}}}

\makeatletter
\long\def\@makecaption#1#2{\vskip\abovecaptionskip{\bf #1.} #2\vskip\belowcaptionskip}
\makeatother

\usepackage[colorlinks,plainpages=false,linkcolor=blue,urlcolor=blue,citecolor=black,pdfpagemode=UseNone,pdfstartview=FitBH]{hyperref}

\begin{document}

\makeatletter
\def\@evenhead{\hfil\itshape\rightmark}
\def\@oddhead{\itshape\leftmark\hfil}
\renewcommand{\@evenfoot}{\hfill\small\raisebox{-1em}{--~\textbf{\thepage}~--}\hfill}
\renewcommand{\@oddfoot}{\hfill\small\raisebox{-1em}{--~\textbf{\thepage}~--}\hfill}
\makeatother

\title[Dispersion and damping of zone-boundary magnons in the noncentrosymmetric superconductor CePt$_3$Si]{\vspace{-5em}\LARGE\\Dispersion and damping of zone-boundary magnons in the noncentrosymmetric superconductor CePt$_3$Si}

\author{D.\,S.~Inosov,$\!^1$ P.~Bourges,$\!^2$ A.~Ivanov,$\!^3$ A.~Prokofiev,$\!^4$ E.~Bauer,$\!^4$ B.~Keimer$^1$}

\address{\begin{tabular}{@{}r@{\,}p{0.845\textwidth}}
$^1$&Max Planck Institute for Solid State Research, Heisenbergstra{\ss}e~1, 70569 Stuttgart, Germany.\\
$^2$&Laboratoire L\'{e}on Brillouin, CEA-CNRS, CEA Saclay, 91191 Gif-sur-Yvette Cedex, France.\\
$^3$&Institut Laue-Langevin, 156X, 38042 Grenoble cedex 9, France.\\
$^4$&Institut für Festkörperphysik, Technische Universität Wien, A-1040 Wien, Austria.\\
\end{tabular}}

\ead{\href{mailto:d.inosov@fkf.mpg.de}{d.inosov@fkf.mpg.de}}

\begin{abstract}
\noindent \hspace*{-1ex}Inelastic neutron scattering (INS) is employed to study damped spin-wave excitations in the non\-centro\-symmetric heavy-fermion superconductor CePt$_3$Si along the antiferromagnetic Brillouin-zone boundary in the low-temperature magnetically ordered state. Measurements along the $(\half\half\kern.5pt L)$ and $(H\,H\,\half\!-\!H)$ reciprocal-space directions reveal deviations in the spin-wave dispersion from the previously reported model. Broad asymmetric shape of the peaks in energy signifies strong spin-wave damping by interactions with the particle-hole continuum. Their energy width exhibits no evident anomalies as a function of momentum along the $(\half\half\kern.5pt L)$ direction, which could be attributed to Fermi-surface nesting effects, implying the absence of pronounced commensurate nesting vectors at the magnetic zone boundary. In agreement with a previous study, we find no signatures of the superconducting transition in the magnetic excitation spectrum, such as a magnetic resonant mode or a superconducting spin gap, either at the magnetic ordering wavevector $(0\,0\kern.5pt\half)$ or at the zone boundary. However, the low superconducting transition temperature in this material still leaves the possibility of such features being weak and therefore hidden below the incoherent background at energies $\lesssim$\,0.1 meV, precluding their detection by INS.
\vspace{-1em}\end{abstract}

\noindent\rule{\textwidth}{.7pt}

\section{Introduction}

Unconventional superconductivity in the noncentrosymmetric heavy-fermion superconductor CePt$_3$Si \cite{BauerHilscher04, BauerKaldarar07} emerges below $T_{\rm c}=0.46$\,K \cite{MukudaNishide09, MicleaMota10} out of an antiferromagnetically (AFM) ordered metallic phase \cite{MetokiKaneko04}. It has been suggested that the lack of inversion symmetry mixes the spin-singlet and spin-triplet superconducting pairing channels, leading to an exotic ground state with an $s$\kern.7pt+\kern.3pt$p$-wave symmetry of the order parameter \cite{TheoryPapers, KlamEinzel09}, supported by recent experimental evidence \cite{YogiKitaoka04, YogiMukuda06, WillersFak09}. An intense discussion about the details of such pairing is under way, in particular concerning the relative magnitude of the singlet and triplet contributions \cite{KlamEinzel09}, and the role of the static AFM order and spin fluctuations for superconductivity \cite{YanaseSigrist07}. A detailed knowledge of the spin excitation spectrum is indispensable to answer these questions.

Recently, in an extensive inelastic neutron scattering (INS) study \cite{FakRaymond08}, it was shown that Kondo-type spin fluctuations are found in the vicinity of the AFM ordering wavevector $(0\kern.5pt0\kern1pt\half)$ in the paramagnetic state \cite{Notation}. Below the N\'{e}el transition temperature, $T_{\rm N}=2.2$\,K, these fluctuations give way to damped spin waves, persisting in a wide range of momenta in the $(H\kern.5pt0\kern.5pt L)$ plane of the reciprocal space. Their dispersion has been successfully parameterized by an effective Heisenberg-type model that involves five principal exchange integrals between localized Ce moments \cite{FakRaymond08}. However, the effects of spin-wave damping via coupling to the continuum of particle-hole excitations across the Fermi level have so far been neglected.

\begin{wrapfigure}[18]{r}{0.32\textwidth}\vspace{-1.4em}
\noindent\includegraphics[width=0.32\textwidth]{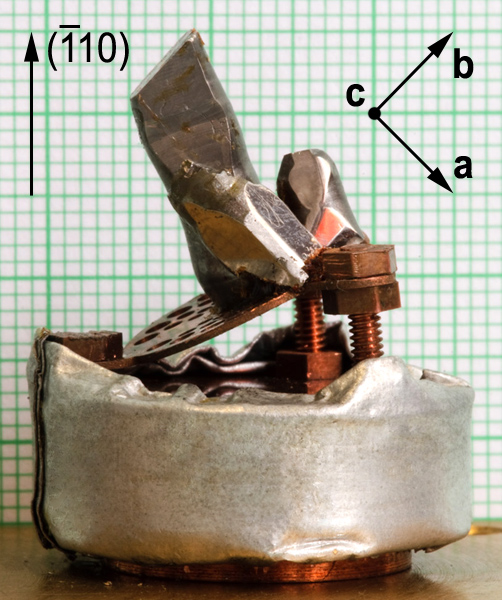}
\caption{Mounting of the single-crystalline CePt$_3$Si ingots for low-temperature INS measurements in the $(H\kern.5pt H\kern.5pt L)$ scat\-tering plane. The copper sample holder (bottom) is shielded with Cd foil.}
\label{Fig:Sample}
\end{wrapfigure}

The Fermi surface of CePt$_3$Si, evaluated both from band structure calculations and quantum-oscillation measurements \cite{HoshimotoYasuda04}, suggests a possible presence of several nesting vectors \cite{KlamManske} that could lead to enhanced spin-wave damping or softening of the spin-excitation energies at certain nearly commensurate wavevectors at the Brillouin zone (BZ) boundary. Such anomalies, if found, could serve as a direct probe of particle-hole scattering and, hence, would shed light on the Fermi surface geometry and electron correlations in this material. This argumentation motivated our present study.

\section{Description of the sample and experimental conditions}

In this paper, we present the results of low-temperature INS measurements performed along the $(\half\half\kern.5pt L)$ direction at the BZ boundary\,---\,a region not covered in previous experiments. We used several single crystals of CePt$_3$Si with a total mass $\sim$\,7\,g that were coaligned using x-ray Laue diffraction and assembled on a copper sample holder in the $(H\kern.5pt H\kern.5pt L)$ scattering plane, as shown in Fig.\,\ref{Fig:Sample}. The overall mosaicity of the sample, measured at half-maximum of the (110) and (002) Bragg peaks, was better than 1$^\circ$. We used a $^3$He dilution insert to cool the sample down to 80\,mK. The triple-axis cold-neutron spectrometer IN14 was operated in the ``W''-geometry in the constant-$k_{\rm f}$ mode, with focusing applied both to the monochromator and analyzer. Second-order neutrons were eliminated by a cold Be-filter placed between the sample and the analyzer. To maximize the intensity, no collimation was used.

\begin{wrapfigure}[15]{r}{0.5\textwidth}\vspace{-1.6em}
\noindent\includegraphics[width=0.5\textwidth]{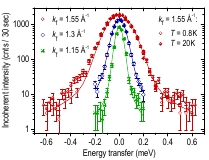}
\caption{Incoherent scattering intensity from the sample for different $k_{\rm f}$, fitted with Voigt profiles.}
\label{Fig:Incoherent}
\end{wrapfigure}

Energy profiles of the incoherent scattering intensity from the sample, measured at $\mathbf{Q}=(0.4~0.4~0)$ for three values of the final neutron momentum, $k_{\rm f}=1.55$\,\AA$^{-1}$, 1.3\,\AA$^{-1}$, and 1.15\,\AA$^{-1}$ (Fig.\,\ref{Fig:Incoherent}), can be well fitted with the Voigt function, yielding full widths of the energy resolution at half-maximum (FWHM) of 0.20 meV, 0.10 meV and 0.077\,meV, respectively. The Lorentzian contribution to the peak width, $w_{\rm L}=0.015$\,meV, was found to be independent of $k_{\rm f}$ within the accuracy of the fit. Its long ``tails'', dominating the incoherent background at low energies, led us to the choice of $k_{\rm f}=1.55$\,\AA$^{-1}$ throughout the experiment as a compromise between energy resolution and the signal-to-noise ratio.

\begin{figure}[t]
\includegraphics[width=\textwidth]{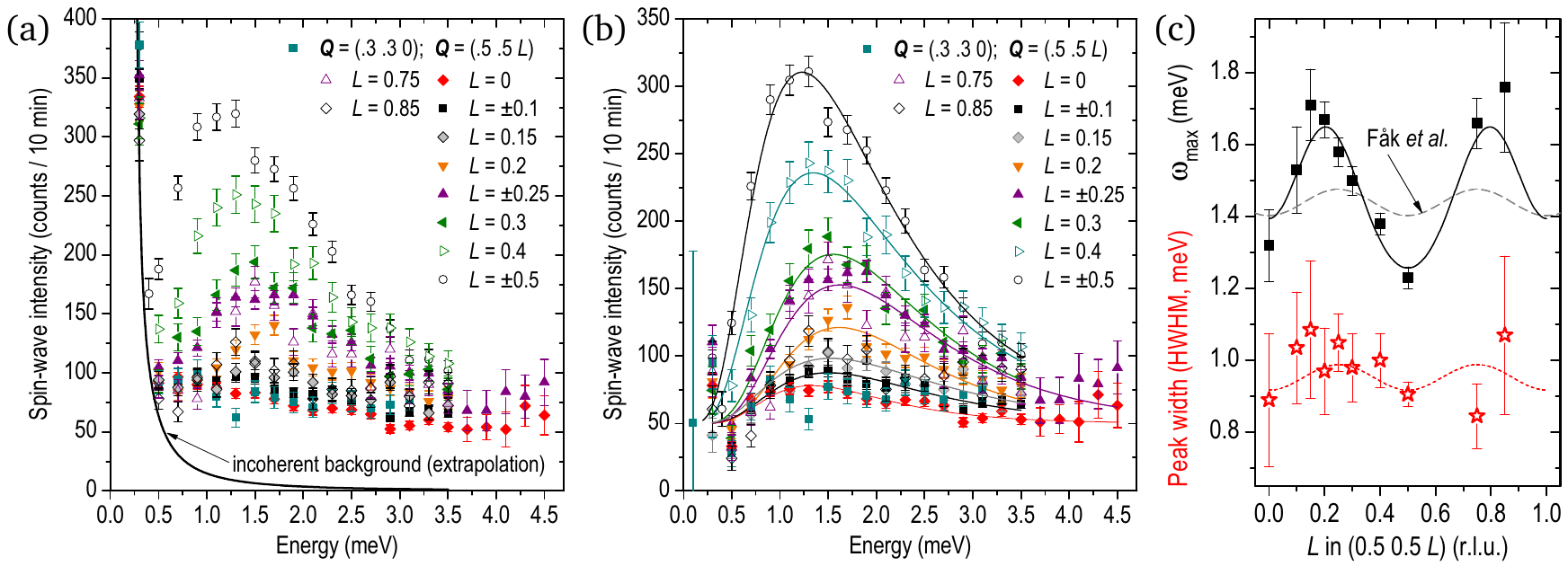}
\caption{Evolution of the magnon intensity along the $(\half\,\half\,L)$ BZ boundary, measured at $T=80$\,mK. (a)~Raw energy scans measured in the magnetically ordered state at different wave vectors, as indicated in the legend. (b)~The same after subtraction of the Voigt-shaped incoherent background. The solid lines result from a global fit to an empirical model. (c)~Dispersion (top) and half-width (bottom) of the peak that resulted from the fit shown in panel (b). The solid line is a guide to the eye. The model of B.~F{\aa}k \textit{et al.} \cite{FakRaymond08} is shown by the dashed line for comparison.\vspace{-1em}}
\label{Fig:Escans}
\end{figure}

\section{Magnons in the magnetically ordered state}

In Fig.\,\ref{Fig:Escans}, we show a series of energy scans measured at different zone-boundary wave vectors $(\half\half\kern.5pt L)$, spanning the irreducible part of the BZ between $L=0$ and $L=\half$. Panel (a) shows the raw data, whereas panel (b) results from a subtraction of the incoherent scattering background, fitted to the Voigt profile (Fig.\,\ref{Fig:Incoherent}), whose extrapolation is given by the solid line in Fig.\,\ref{Fig:Escans}\,(a). The strongly asymmetric lineshape of the resulting magnetic signal could not be well fitted by the damped harmonic oscillator model (possibly due to a resolution effect). Therefore, an empirical fit shown by solid lines in Fig.\,\ref{Fig:Escans}\,(b), which includes a constant background offset shared by all curves, was used. The resulting magnon dispersion (defined here by the positions of the peak maxima) and the half width at half maximum (HWHM) of the peak, related to the magnon lifetime, are shown in Fig.\,\ref{Fig:Escans}\,(c). Here, the solid line is given by the fit of the experimental dispersion to a sum of sinusoidal functions, whereas the dashed black line shows the dispersion given by the spin-wave model of B.~F{\aa}k \textit{et al.} \cite{FakRaymond08} along the same reciprocal-space direction. One can see that the latter is characterized by a somewhat smaller amplitude as compared to the directly measured one.


\hvFloat[floatPos=t, capWidth=1.0, capPos=r, capVPos=t, objectAngle=0]{figure}
        {\includegraphics[width=0.6\textwidth]{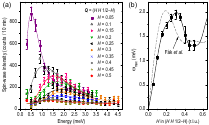}~~~~}
        {Magnon dispersion along the $(H\,H\,\half\!-\!H)$ direction, measured at $T=0.8$~K. (a)~Energy scans after subtraction of the Voigt-shaped incoherent background. (b)~Dispersion of the peak maximum. The spin-wave model of B.~F{\aa}k \textit{et al.} \cite{FakRaymond08} is shown by the dashed line for comparison. The solid lines are guides to the eyes.}
        {Fig:EscansHHL}

A similar analysis of the magnon dispersion is presented in Fig.\,\ref{Fig:EscansHHL} for the $(H\,H\,\half\!-\!H)$ direction, which connects the AFM ordering wavevector $(0\,0\,\half)$ with the zone boundary at $(\half\,\half\,0)$. These data were measured at $T=0.8\,\mathrm{K}>\!T_{\rm c}$, but since the spin-wave spectrum is insensitive to the SC transition, these experimental conditions are practically equivalent to those in Fig.\,\ref{Fig:Escans}. A fit of the experimental dispersion, shown in panel (b) by a solid line, exhibits a pronounced nonmonotonic behavior that deviates from the spin-wave model of Ref.\,\citenum{FakRaymond08} (dashed line) by up to 20\%.

\begin{wrapfigure}[21]{r}{0.45\textwidth}\vspace{-0.2em}
\noindent\includegraphics[width=0.45\textwidth]{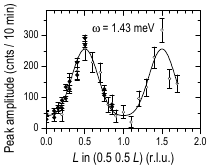}
\caption{$L$-dependence of the background-subtracted intensity at the peak maximum along $(\half\,\half\,L)$, modulated due to the AFM structure factor. Empty symbols represent the directly measured intensity at $\omega=1.43$\,meV, whereas solid symbols result from the amplitudes of the fits presented in Fig.\,\ref{Fig:Escans}\,(b). The least-squares fit (solid line) is given by the following periodic function: $I(L)=I_0\,[1-0.827\cos2\piup L+0.161\cos 4\piup L].$}
\label{Fig:StructureFactor}
\end{wrapfigure}

The strong variation in measured inelastic intensity between $L=0$ and $L=1/2$ in both figures results from the momentum-dependent AFM structure factor, which is maximized at the AFM ordering wavevector $(0\,0\,\half)$ and becomes strongly reduced, but not vanishing, at $(\half\,\half\,0)$. This periodic intensity modulation is shown in Fig.\,\ref{Fig:StructureFactor} after subtraction of the constant background, and can be approximated with a sum of the first- and second-harmonic sinusoidal functions of $L$ (solid line).

\begin{figure}[t]
\includegraphics[width=\textwidth]{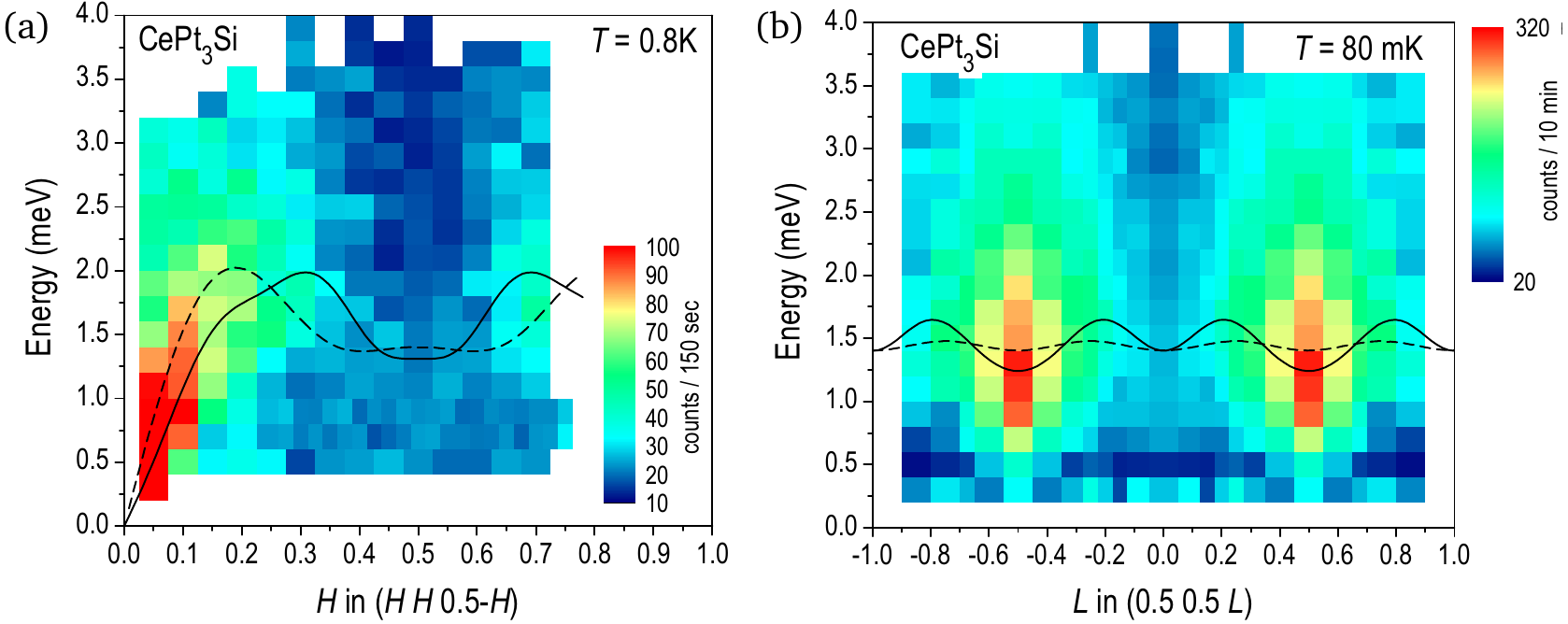}
\caption{Color maps of the magnon intensity along the $(H\,H\,\half\!-\!H)$ and $(\half\,\half\,L)$ reciprocal-space directions. In panel (b), symmetrization with respect to the $L=0$ line has been applied. The solid and dashed lines represent fits of the experimental dispersion, defined as the energy of the peak maximum at every momentum, and the spin-wave model of B.~F{\aa}k \textit{et al.} \cite{FakRaymond08}, respectively.\vspace{-1em}}
\label{Fig:SpinWaves}
\end{figure}

The same structure-factor modulation can also be seen in the color maps of the INS intensity that are shown in Fig.\,\ref{Fig:SpinWaves} (a) and (b) for the $(H\,H\,\half\!-\!H)$ and $(\half\,\half\,L)$ reciprocal-space directions, respectively. The fits of the experimental dispersion (same as in Fig.\,\ref{Fig:Escans} and \ref{Fig:EscansHHL}) are summarized here by solid lines and compared to the model of B.~F{\aa}k \textit{et al.} \cite{FakRaymond08} shown in dashed lines. In spite of the apparent differences between these two fits in both directions, the disparity always remains smaller than the  intrinsic energy width of the overdamped signal. These color maps also demonstrate the absence of any pronounced additional branches of paramagnon excitations in the vicinity of $(\half\,\half\,0)$ that could originate from the nested sections of the normal-state Fermi surface.

\section{Superconducting state}

\begin{figure}[b]
\includegraphics[width=\textwidth]{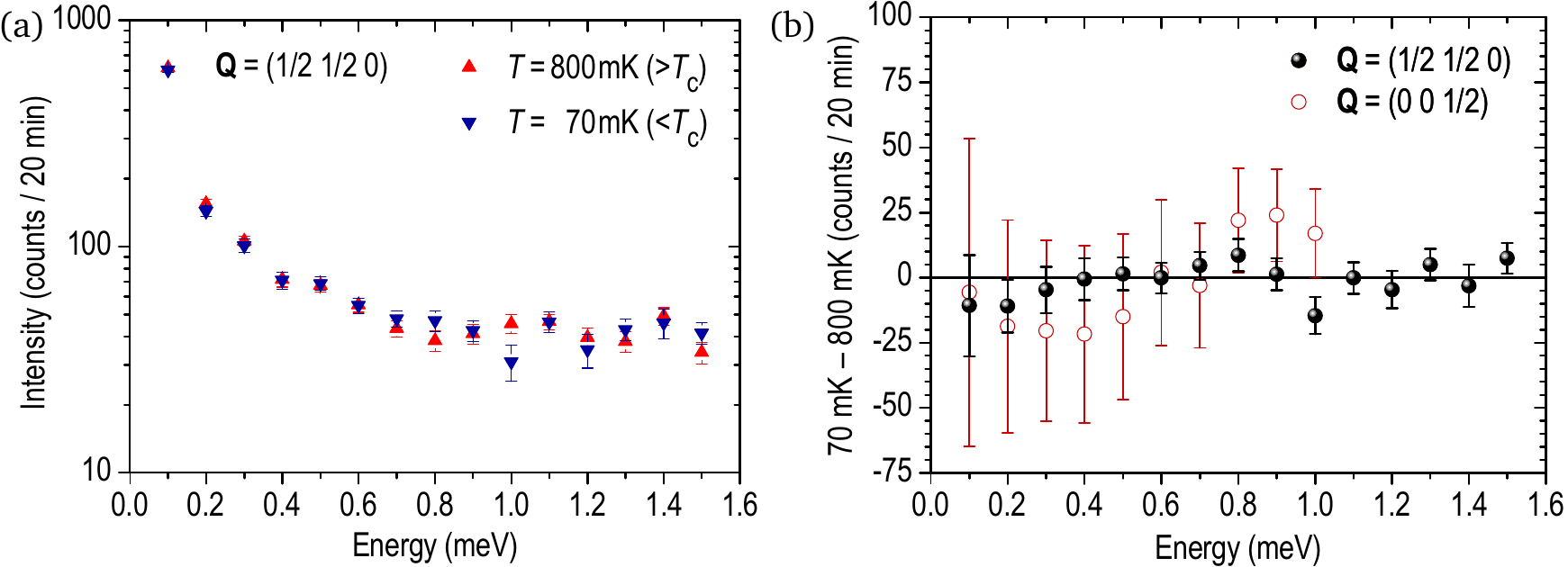}
\caption{(a) Raw energy scans at $(\half\,\half\,0)$ above and below $T_{\rm c}$. (b) Subtraction of the two data sets (black points) reveals no spin-resonance enhancement within the experimental error. The respective temperature subtraction at the AFM propagation vector $(0\,0\,\half)$ is shown by empty symbols.}
\label{Fig:Tdifference}
\end{figure}

In the unconventional heavy-fermion superconductor CeCoIn$_5$ ($T_{\rm c}=2.3$\,K), a pronounced magnetic resonant mode, centered at $\sim$\,0.6\,meV, emerges below $T_{\rm c}$ out of a weak and featureless normal-state spectrum \cite{StockBroholm08}. This resonance is similar to the one found in high-$T_{\rm c}$ cuprates \cite{RossatMignod91} and iron pnictides \cite{InosovPark10}, where it serves as the hallmark of a sign-changing superconducting order parameter. However, no such resonant enhancement has been reported neither in the spin-triplet $p$-wave superconductor Sr$_2$RuO$_4$ \cite{BradenSidis02} nor in the earlier experiments on CePt$_3$Si \cite{FakRaymond08} close to the magnetic ordering wave vector. We have therefore performed a comparison of the inelastic signal in CePt$_3$Si above and below $T_{\rm c}$ to establish the absence of resonant effects both near the zone boundary at $(\half\,\half\,0)$ and at the AFM propagation vector, $(0\,0\,\half)$. This comparison is presented in Fig.\,\ref{Fig:Tdifference}. Indeed, within the statistical error, the inelastic intensity remains unchanged between 800\,mK ($>T_{\rm c}$) and 70\,mK ($<T_{\rm c}$), in agreement with a previous report \cite{FakRaymond08}. This can be best seen in the difference spectra in Fig.\,\ref{Fig:Tdifference}\,(b), which are indistinguishable from zero within the statistical uncertainty for both high-symmetry points. The larger error bars of the $\mathbf{Q}=(0\,0\,\half)$ data points are explained by the intense ``tails'' of the magnetic Bragg peak that persist at this AFM wave vector in the low-energy region. This observation can not exclude a weak magnetic resonance at extremely low energies below 0.1\,meV, which is plausible taking into account that $T_{\rm c}$ of CePt$_3$Si is a factor of six lower than in CeCoIn$_5$. Such a scenario would render it practically unobservable to INS due to the high level of incoherent-scattering background at these low energies. We also can not exclude the possibility of a resonant mode located at an incommensurate wave vector not covered by the present and previous studies, although this alternative would be rather unusual.

\section{Discussion}

In conclusion, our experiments revealed no additional branches of paramagnetic excitations along the $(\half\half\kern.5pt L)$ direction, apart from the strongly damped spin-wave modes, and found a nearly constant intrinsic line width of the spin-wave excitations without pronounced anomalies. These observations point to the absence of strong nesting vectors in the Fermi surface of CePt$_3$Si along the zone boundary.

The strong damping of spin-wave excitations in the magnetically ordered state results in peak widths of the order of the spin-wave energies, leading to the failure of the conventional damped-oscillator model to describe the spectral line shape appropriately. This questions the applicability of localized Heisenberg-type models for the description of the spin-wave spectrum in this metallic system, as in such a case the spin-wave energies are no longer well defined except for the immediate vicinity of the magnetic Bragg reflections. Deviations of up to 20\% in the dispersion of the peak maximum from the previously constructed Heisenberg-type model are found, but this disparity remains smaller than the intrinsic energy width of the peak.

We also found no anomalies in the INS signal across the superconducting transition, which essentially excludes the possibility of an intense commensurate resonant mode at the AFM wave vector, similar to the one found in CeCoIn$_5$ \cite{StockBroholm08}. Yet, a weak resonance at an extremely low energy or at an incommensurate $Q$ vector would still be consistent with our results.

\section*{Acknowledgements}

The authors acknowledge stimulating discussions and support from V.~Hinkov, T.~Keller, L.~Klam, Yuan~Li, D.~Manske, C.~Pfleiderer and thank M.~Ohl for the original design of the sample holder. The INS experiment has been conducted at the Institut Laue-Langevin (ILL) in Grenoble. Some preliminary data have also been collected at the 4F2 spectrometer at the Laboratoire L\'{e}on Brillouin, Saclay, France. Parts of this work have been supported by the Austrian FWF P22295.


\begin{thebibliography}{10}
\bibitem{Notation} Here and in the following, reciprocal-space vectors $(H\kern.5pt K L)$ are given in dimensionless reciprocal lattice units (r.l.u.), i.e. in units of the reciprocal-lattice vectors $2\pi/a$, $2\pi/b$, and $2\pi/c$, where $a=b=4.072$\,\AA\ and $c=5.450$\,\AA\ are low-temperature lattice constants, as determined during sample alignment.
\end{thebibliography}
\end{document}